\documentstyle[12pt,epsfig]{article}
\renewcommand{\theequation}{\arabic{equation}}
\renewcommand{\thesection}{\arabic{section}}
\newcommand{\bea}{\begin{eqnarray}}
\newcommand{\ena}{\end{eqnarray}}
\newcommand{\vs}[1]{\vspace{#1 mm}}

\newcommand{\tp}{{\tilde {z}}}

\begin{document}
\noindent
\topmargin 0pt
\oddsidemargin 5mm

\begin{titlepage}
\setcounter{page}{0}
\thispagestyle{empty}
\begin{flushright}
June 8, 2001\\
OU-HET 389\\
hep-ph/0106086\\
\end{flushright}
\vs{4}
\begin{center}
{\LARGE{\bf The matter fluctuation effect to T violation}} \\ 
\vs{2}
{\LARGE{\bf at a neutrino factory}}\\
\vs{6}
{\large 
Takahiro Miura\footnote{e-mail address:
miura@het.phys.sci.osaka-u.ac.jp},
Tetsuo Shindou\footnote{e-mail address:
shindou@het.phys.sci.osaka-u.ac.jp},
Eiichi Takasugi\footnote{e-mail address:
takasugi@het.phys.sci.osaka-u.ac.jp},\\
and Masaki Yoshimura\footnote{e-mail address:
masaki@het.phys.sci.osaka-u.ac.jp}
\\
\vs{2}
{\em Department of Physics,
Osaka University \\ Toyonaka, Osaka 560-0043, Japan} \\
}
\end{center}
\vs{6}
\centerline{{\bf Abstract}}
We derived an analytic formula for T violation by using 
the perturbation theory for small quantities, 
$\Delta m_{21}^2L/2E$ and $\delta a(x)L/2E$, where $\delta a(x)$ 
represents symmetric and asymmetric matter fluctuations, i.e., 
deviations from the average density.  
We analyzed the effect of matter fluctuations 
to T violation, by assuming PREM profile of earth matter density. 
We found that matter fluctuations do not give any 
viable contribution for $L<$6000km, while the fluctuation effect 
becomes large due to resonances for $L>$7000km.  
For 7000km$<L<$8000km, matter fluctuations contribute  
destructively to the average density term and 
the net result is small, while  for $L>$8000km, the contribution 
from matter fluctuations becomes large but contributes 
constructively. 

\end{titlepage}

\newpage

\section{Introduction}

In our previous paper[1], we analyzed the matter 
fluctuation effect to T violation, 
$P(\nu_\alpha \to \nu_\beta)-P(\nu_\beta \to \nu_\alpha)$ 
$(\alpha \neq \beta)$ at a neutrino factory[2]
by using the perturbation method developed by Koike 
and Sato[3] and Ota and Sato[4]. The perturbation is made 
with respect to small quantities, $\Delta m_{21}^2L/2E$ and 
$\delta a(x)L/2E$, where $\delta a(x)$ represents matter 
fluctuations, i.e., deviations from the average density . 
We examined T violation up to the 2nd order and found that 
it arises from the average density, 
the 1st order term which is proportional to 
$\Delta m_{21}^2L/2E$ and term from matter fluctuations, the 2nd 
order term proportional to $(\Delta m_{21}^2L/2E)(\delta a(x)L/2E)$. 
The zeroth order term and terms proportional to  
$(\Delta m_{21}^2L/2E)^2$ and $(\delta a(x)L/2E)^n$ $(n=1,2,3)$ 
do not contribute to T violation. 
The 1st order term and the 2nd order term from symmetric 
matter fluctuations which we denote $\delta a_s$ contribute 
to $\sin \delta$ 
term and the 2nd order term from asymmetric matter fluctuations, 
$\delta a_a$ does to the fake 
$\cos \delta$ term, where $\delta$ is the CP violation phase 
in MNS neutrino mixing matrix[5]. 

By using the preliminary reference earth model (PREM)[6] 
for symmetric matter fluctuations and assuming that 
asymmetric matter fluctuations are much less than 
symmetric matter fluctuations given by PREM, we 
computed T violation and found that 
the 2nd order term from symmetric and asymmetric matter 
fluctuations gives only negligible contributions to 
T violation, and thus the constant (average) matter 
approximation is valid for $L=3000$km. On the other hand, 
for $L=7332$km, we found that 
the contribution from symmetric matter fluctuations 
becomes as large as the 1st order term and 
moreover they contribute destructively  so that T violation 
becomes very small. This means that the constant (average) 
matter approximation is not valid for $L>$7000km and also 
the validity of our 2nd order formula should be examined. 

In this paper, we discuss the following three questions: 
(1) Is the 2nd order formula valid? 
(2) What is the length where the constant (average) matter 
approximation fails for T violation? That is, 
at what length, the matter fluctuation effect becomes 
important. (3) What is the size of T violation 
for $L>$7000km? 

To answer these questions, we computed the next order 
(3rd order) contribution to T violation, i.e., the 
term proportional to $(\Delta m_{21}^2L/2E)
(\delta a(x)_s L/2E)^2$. Our result is as follows: 
The contribution from matter fluctuation can be safely 
neglected for $L<$6000km. When we discuss T violation 
with length larger than 6000km, the matter fluctuation 
effect should be taken into account. The 3rd order 
contribution is negligible in comparison with the 1st and 
the 2nd order term for all distances. Therefore, 
T violation can be reliably estimated for all distances and 
it becomes very small for 7000km$<L<$8000km. For $L>$8000km, 
the 1st and the 2nd terms contribute constructively and 
T violation becomes large.

This paper is organized as follows: In Sec.2, the analytic 
formula for T violation is given. The numerical analysis 
for T violation by using PREM profile is given in Sec.3 
and the mechanism how the cancellation occurs for 
7000km$<L<$8000km is explained analytically. The summary is 
given in Sec.4. The derivation of the analytic formula is 
given in Appendix.

\section{T violation formula}

For completeness, we give the 1st and the 2nd order contributions 
to T violation and the definition of parameters in the formula. 
We also give the 3rd order contribution from symmetric fluctuations.

\noindent
(a) Notation

We begin with defining the neutrino mixing matrix as
\bea
U
&=&e^{i\theta_{y} \lambda_7}{\rm diag }(1,1,e^{i\delta})
 e^{i\theta_{z} \lambda_5}e^{i\theta_{x}\lambda_2}
\nonumber\\
&=&\pmatrix{
c_{x} c_{z} &s_{x} c_{z}  & s_{z} \cr
-s_{x} c_{y} -c_{x} s_{y} s_{z} e^{i\delta} &
c_{x} c_{y} -s_{x} s_{y} s_{z} e^{i\delta} &
s_{y} c_{z} e^{i\delta}\cr 
s_{x} s_{y}-c_{x} c_{y} s_{z} e^{i\delta} &
-c_{x} s_{y}-s_{x} c_{y} s_{z} e^{i\delta} &
c_{y} c_{z}e^{i\delta}\cr}\;,
\ena
where $\lambda_j$ ($j=2,5,7$) are Gell-Mann matrices and 
$c_{a}=\cos \theta_{a}$ and $s_{a}=\sin \theta_{a}$. 
Since the Majorana CP-violation phases are irrelevant to the 
neutrino oscillations (flavor oscillations)[7], we neglected them. 
The relation between the flavor eigenstates, 
$|\nu_\alpha\rangle$ $(\alpha=e,\mu,\tau)$, 
and the mass eigenstates, $|\nu_i\rangle$ $(i=1,2,3)$, is given by
\bea
|\nu_\alpha\rangle=U_{\alpha i}|\nu_i\rangle\;.
\ena

The evolution of the flavor eigenstates in matter with energy $E$ 
is given by
\bea
i\frac{d}{dx}|\nu_\beta(x)\rangle=
H(x)_{\beta \alpha}|\nu_\alpha(x)\rangle\;,
\ena
where Hamiltonian $H(x)_{\beta \alpha}$ is given by 
\bea
H(x)_{\beta \alpha}
=\frac{1}{2E}\left\{U_{\beta i}
\pmatrix{0&&\cr &\Delta m_{21}^2 &\cr 
&&\Delta m_{31}^2}_{ii}U_{i\alpha}^\dagger+
\pmatrix{a(x)&&\cr &0& \cr &&0}_{\beta\alpha}\right\}\;.
\ena
Here $\Delta m_{ij}^2 \equiv m_i^2-m_j^2$ with $m_i$ being 
the mass of $|\nu_i\rangle$, $G_F$ is the
Fermi coupling constant and  
\bea
a(x)\equiv 2\sqrt{2}G_F n_e(x) E=7.56\times 10^{-5}
\left(\frac{\rho(x)}{\rm g/cm^3}\right)
\left(\frac{Y_e}{0.5}\right)
\left(\frac{E}{\rm GeV}\right){\rm eV^2}\;,
\ena
where $n_e(x)$, $Y_e$ and $\rho(x)$ are 
the electron number density, the electron fraction 
and the matter density, respectively. For the electron 
fraction, we use $Y_e=0.5$.

We separate the matter density fluctuation from its average $\bar a$, 
\bea
\delta a(x) \equiv a(x)-\bar a\;,
\ena
and consider the deviation $\delta a(x)$ as a perturbative 
term. That is, we solve the evolution equation by treating 
$\delta a(x)L/2E$ and $\Delta m_{21}^2L/2E$ as perturbative 
terms, because they are small for most of the cases of 
planned neutrino factories.

T violation is defined by
\bea
\Delta P_{\nu_\alpha \nu_\beta}^{T}
=P(\nu_\alpha \to \nu_\beta)-P(\nu_\beta \to \nu_\alpha)\;,
\ena
which is evaluated by using the method developed by 
Koike and Sato[3], Ota and Sato[4]. Ota and Sato showed that 
the 1st order approximation for $\delta a(x)L/2E$ is good 
enough to reproduce the transition probability. However, 
T violation is as small as a few $\%$ of the transition 
probability so that this approximation is not valid. In 
fact, we showed the 2nd order term becomes as important as
the 1st order term for $L>$6000km. We show T violation, 
$\Delta P_{\nu_e \nu_\mu}^T$ up to the 3rd order perturbation 
for symmetric matter fluctuations and the 2nd order for 
asymmetric matter fluctuations. The analytical formula 
is given by expanding symmetric and asymmetric matter 
fluctuations in terms of Fourier series as 
\bea
\delta a(x)_s = \sum_{n=\pm 1,\pm 2,\cdots} a_{2n}
e^{-iq_{2n}x}\;,\quad
\delta a(x)_a = \sum_{n=0,\pm 1,\cdots} a_{2n+1} e^{-iq_{2n+1}x}\;,
\ena
where
\bea
q_{n}=\frac{n\pi}{L}\;.
\ena

\noindent
(b) The analytic formula for T violation

In order to define T violation, we define the following 
quantities:
\bea
\tan 2 \theta_{\tilde z}
&=&\frac{s_{2z}(\Delta m_{31}^2-\Delta m_{21}^2 s_{x}^2)}
        {c_{2z}(\Delta m_{31}^2-\Delta m_{21}^2 s_{x}^2)- \bar
a}\;,\nonumber\\
\lambda_\pm
&=&\frac{1}{2}
        \left(\Delta m_{31}^2+\Delta m_{21}^2 s_{x}^2+\bar a\right.\;
\nonumber\\
&&\hskip 1mm \left. \pm \sqrt{
        \{c_{2 z}(\Delta m_{31}^2-\Delta m_{21}^2 s_{x}^2)-\bar a\}^2
        +s_{2z}^2(\Delta m_{31}^2-\Delta m_{21}^2 s_{x}^2)^2}
        \right)\;,\nonumber\\
k_1&=&\frac{ \Delta m^2_{21}c_x^2 - \lambda_- }{2E}\;,\nonumber\\
k_2&=&\frac{ \lambda_+ - \Delta m^2_{21}c_x^2 }{2E}\;,\nonumber\\
k&=&\frac{ \lambda_+ - \lambda_- }{2E}\;.
\ena

The sum of the 1st and the 2nd order terms due to symmetric matter 
fluctuations contributes 
to $\sin \delta$ term and is given 
by
\bea
\Delta P_{\nu_e \nu_\mu}^{T(1+2s)}
&=&-\frac{\Delta m^2_{21}}{E}s_{2x}s_{2y}s_{2\tp}s_\delta 
\left[\frac{c_{\tp}c_{z-\tp}}{k_1}+\frac{s_{\tp}s_{z-\tp}}{k_2}
\right]\sin \frac{k_1L}2 
\sin \frac{k_2L}2 \sin \frac{kL}2
\nonumber\\
&\times&\left[1+2
\sum_{n=1,2,\cdots}\frac{a_{2n}}{2E}
\left(\frac{c_\tp^2 k_1}{k_1^2-q_{2n}^2}
 -\frac{s_\tp^2 k_2}{k_2^2-q_{2n}^2}
 +\frac{c_{2\tp}k}{k^2-q_{2n}^2}
\right)\right]
\;.
\nonumber\\
\ena
The asymmetric matter fluctuation contributes to 
the $\cos \delta$ term and is given by
\bea
\Delta P_{\nu_e \nu_\mu}^{T(2a)}
&=&-\frac{\Delta m^2_{21}}{E}s_{2x}s_{2y}s_{2\tp}c_\delta 
\left[\frac{c_{\tp}c_{z-\tp}}{k_1}+\frac{s_{\tp}s_{z-\tp}}{k_2}
\right]\sin \frac{k_1L}2 \sin \frac{k_2L}2 \sin \frac{kL}2
\nonumber\\
&\times&
\left\{\sum_{n=0,1,\cdots}\frac{a_{2n+1}}{2E} 
\frac{k}{k^2-q_{2n+1}^2}\cot \frac{kL}2
-\frac{c_{\tp}^2k_1}{k_1^2-q_{2n+1}^2}\cot \frac{k_1 L}2
\right.
\nonumber\\
&&\left.
-\frac{s_{\tp}^2k_2}{k_2^2-q_{2n+1}^2}\cot \frac{k_2 L}2\right\}
\;.
\ena

The 3rd order contribution from symmetric fluctuations is 
given by 
\bea
\Delta P_{\nu_e \nu_\mu}^{T(3s)}
&=&-\frac{\Delta m^2_{21}}{E}s_{2x}s_{2y}s_{2\tp}s_\delta 
\left[\frac{c_{\tp}c_{z-\tp}}{k_1}+\frac{s_{\tp}s_{z-\tp}}{k_2}
\right]
\nonumber\\
&&\left (\left\{
\sum_{n,m=\pm 1,\pm 2,\cdots}\frac{a_{2n}a_{2m}}{(2E)^2}
\left [\frac{c_{2\tilde z}}{k+q_{2n}}
  \left (\frac{c_{\tilde z}^2}{k_1+q_{2m}}-
  \frac{ s_{\tilde z}^2}{k_2-q_{2m}}\right)
  \right.\right.\right.
  \nonumber\\
&&+\left(\frac{c_{\tilde z}^2}{k_1+q_{2n}}-
  \frac{ s_{\tilde z}^2}{k_2-q_{2n}}\right)
  \left(\frac{c_{\tilde z}^2}{k_1+q_{2n+2m}}-
  \frac{ s_{\tilde z}^2}{k_2-q_{2n+2m}}\right)
  \nonumber\\
  &&  \left.
  +\frac{1}{k+q_{2n}}
  \left(\frac{c_{\tilde z}^2}{k_1+q_{2n+2m}}-\frac12
  \frac{ s_{\tilde z}^2}{k_2-q_{2n}}\right)   
  \right\} 
  \sin \frac{k_1L}2 \sin \frac{k_2L}2 \sin \frac{kL}2
\nonumber\\
&&\left.
+\frac{s_{2\tilde z}^2}4\left(\sum_{n=\pm 1,\pm 2,\cdots}
\frac{a_{2n}a_{-2n}}{(2E)^2}\frac{L}{k-q_{2n}}\right)
\left(\sin \frac{k_1L}2 
\sin \frac{k_2L}2 \cos \frac{kL}2+\frac12 \sin^2\frac{kL}2
\right)\right )\;.
\nonumber\\
\ena

It is amusing to see that the 1st and the 2nd order terms due to 
symmetric matter fluctuations  have the same coefficient 
and the same oscillation term as we can see from Eq.(11). 
If one of resonance conditions, $k_1^2=q_2^2$, $k_2^2=q_2^2$ 
and $k^2=q_2^2$ is realized at some distance, the matter 
fluctuation term dominates over the 1st order term, although 
this singular behavior is cancelled by the oscillation term. 
The asymmetric matter contribution gives the similar 
contribution except for the angle $\delta$. 
The 3rd order term is rather complicated and 
seems to have double poles when $n=m$, which is false 
due to the cancellation between other terms, and there are
no singularities. 

As we expected, symmetric and asymmetric matter fluctuations 
contribute to the $\sin \delta$ and $\cos \delta$ parts of 
T violation.

The important fact for T violation is that the  
relation 
\bea
\Delta P_{\nu_e \nu_\mu}^{T}=
\Delta P_{\nu_\mu \nu_\tau}^{T}=
\Delta P_{\nu_\tau \nu_e}^{T}\; ,
\ena
holds even in our 3rd order formula. That is, we can not 
gain any further information by examining other channels. 
This fact for the constant matter was first found by 
Krastev and Petcov[8] and the validity for the 2nd order 
formula is proved in Ref.1.

\section{Numerical analysis}

By using the analytic formula, 
$\Delta P_{\nu_e \nu_\mu}^{T(1+2s+2a+3s)}$, we investigate 
the $L$ and $E$ dependences of T violation. The energy of 
neutrino, $E$ and the distance, $L$ are taken from 
1GeV to 30GeV and from 1000km to 12000km. 
Neutrino oscillation parameters are taken as 
\bea
\Delta m^2_{31}&=&3\times 10^{-3} \;({\rm eV^2})\;,\quad 
\Delta m^2_{21}=5\times 10^{-5} \;({\rm eV^2})\;,\nonumber\\
\sin 2\theta_x &=& \sin 2\theta_y = 1\;,\quad 
\sin \theta_z = 0.1\;,\quad \delta = \pi/4\; ,
\ena
as a typical case.

\vskip 2mm
\noindent
(a) General features of T violation

Here we examine the effect of symmetric fluctuations to 
T violation. For the average density and symmetric 
fluctuations, we assume PREM and decompose it 
into the average density $\bar{a}$ and the 
Fourier coefficients $a_{2n}$ ($n=1,2,3,4$), which are 
functions of $L$. It is a general 
belief that the average density is determined with better 
accuracy than matter fluctuations.

We  computed T violation, $\Delta P_{\nu_e\nu_\mu}^{T}$ 
from $L=$1000km to 12000km. We found that the 2nd and 
the 3rd order terms are not important for $L<6000$km and 
thus the constant (average) matter approximation is valid. 
We remind that the 1st order term contains the effect of 
the constant matter, while the 2nd and the 3rd order 
terms are due to matter fluctuations as we can see in Eqs.(11), 
(12) and (13). 

Now we concentrate on the distance $L>6000$km.  
In Figs.1-6, we show the $E$ dependence of T violation 
for $L=$6000km, 7000km, 7700km, 8000km, 10000km and 
11000km, where the 1st, the 2nd and the 3rd 
order terms are shown by the dashed line, the dotted line 
and the dash-dotted line, respectively. The solid line 
represents the sum of them. The vacuum (no matter) case 
is shown by the dash-twodotted line for comparison. 

When $L>$7000km, the 2nd order term becomes 
the same order as the 1st order term and is no more 
neglected. Moreover, as we see in Figs.2 and 3, 
for 7000km$<L<$8000km, 
the 1st and the 2nd order terms contribute 
destructively and T violation becomes small.  
When 8000km$<L<$9000km, the 1st and the 2nd order 
terms contribute constructively as we see in Fig.4. 
For $L>$10000km, the 2nd order term becomes much larger than the 
1st order term. 

For all distances, the 3rd order term 
(the dash-dotted line) gives only a negligible contribution 
and it is hard to see from these figures. Thus, the 3rd 
order term can safely be neglected so that we need to 
consider the 1st and the 2nd order terms only. 

When we see Figs.1-6, for $E>5$GeV, we observe that  
there are two peaks. One is at $E=$5.5GeV and the 
other is at $E=$10GeV for $L=$6000km. The energies 
of these peaks increase as $L$ increases. T violation 
for energies larger than that of the lower energy peak 
lies between these two peak values. 

We examined the $L$ dependence of these two peak 
values and the result is shown in Fig.7a. Faint diamonds 
and dark diamonds correspond to the peak values of 
the lower and the higher energy peaks, respectively. 
Diamonds,  stars, boxes, triangles, 
circle show T violation for $s_z=\sin \theta_{13}=$ 
0.1, 0.08, 0.06, 0.04 and 0.02. 
Fig.7b shows the case for $E=$30GeV. 

Due to the cancellation between the 1st and the 2nd order 
terms, T violation becomes zero  
at around 7300km. For $E=$30GeV, there are two zeros at 
around $L=$7300km and 9600km. 

Another aspect we can observe from these figures is that
the $s_z$ dependences of T violation is roughly linear.  

\vskip 2mm
\noindent
(b) The relative sign between the 1st and the 2nd order terms

In order to see the changes of the relative sign 
between the 1st and the 
2nd order terms, we write  these terms omitting the overall 
factor and the oscillation term as 
\bea
1+2\sum_{n=1,2,\cdots}\frac{a_{2n}}{2E}
\left(\frac{c_\tp^2 k_1}{k_1^2-q_{2n}^2}
 -\frac{s_\tp^2 k_2}{k_2^2-q_{2n}^2}
 +\frac{c_{2\tp}k}{k^2-q_{2n}^2}
\right)\; .
\ena
The 1st order term is expressed by 1, while the 2nd order term 
is expressed by the sum of Fourier coefficients, $a_{2n}$. 
The singularities at $k_1^2=q_{2n}^2$,  $k_2^2=q_{2n}^2$ 
and $k^2=q_{2n}^2$ correspond to resonances. In general, 
the 2nd order term is smaller than the 1st order term if 
$k_1$, $k_2$ and $k$ are away from resonance points. 
Since $q_{2n}=2n\pi/L$, we compared $(k_1L/2)^2$, 
$(k_2L/2)^2$ and $(kL/2)^2$ with $(n\pi)^2$ for $E>$5GeV 
in Fig.8. 
For $L$ is as small as 3000km, the resonance condition 
is not satisfied so that the 2nd term gives only 
a negligible contribution. 

We consider the change of the relative sign between 
the 1st and the 2nd order terms as $L$ increases 
from 7000km to 8000km at the lower energy  peak 
position. Since $a_2<0$, $k_1<0$, $k_2>0$ and $k>0$, 
the sign of the $k_1$ term is negative for $L$=7000km and 
becomes positive at $L=$8000km. The resonance occurs 
at around $L=$7700km. The same is true for $k_2$ term. 
The $k$ term never reaches to the resonance point. 
That is, when $L$ is as small as 6000km, all terms 
can be neglected because they never approach to 
the resonance point. As a result, T violation is 
positive. 
When $L$ exceeds 7000km,  
contributions from the $k_1$ and $k_2$ terms become 
important and their signs are negative. As $L$ 
approaches to 7700km, their contribution cancelles 
the 1st order term and T violation vanishes at around 7550km. 
Between 7550km$<L<$7700km, both $k_1$ and $k_2$ terms 
become still negative and dominates over the 1st term 
so that T violation becomes negative. At the resonance 
point, this singularity is cancelled by the oscillation 
term and T violation obtains non-zero negative value. 
After passing the resonance point, both $k_1$ and $k_2$ 
terms become positive and contribute additively to the 
1st order term. Since the 
oscillation term becomes negative and thus T violation 
remains to be negative. 

The $L$ dependence of T violation at the larger energy 
can be understood similarly. For $E=$30GeV, the zero of 
T violation at around 7300km can be understood similarly. 
The zero for 9600km is due to the resonance for $k$. 

\vskip 2mm
\noindent
(c)The effect from the uncertainty for the average matter density

So far, we used PREM to derive the average density and 
the matter fluctuations. For the sake of argument, we consider 
5$\%$ uncertainty for the average matter density although 
we do expect that the uncertainty is much less. 
We examined how T violation changes when the average density is 
changed by 5$\%$. If the average matter density changes, 
it affects to the angle $s_\tp$ and $k_1$, $k_2$ and $k$ as 
we can see in Eqs.(10). As a result, the distance $L$ 
where the resonance occurs changes. In Fig.9, 
we show the $L$ dependence of T violation. Diamonds, boxes and 
stars correspond to the average value from PREM, $5\%$ smaller 
value and $5\%$ larger value. Faint and 
dark ones correspond the lower energy peak and the higher energy peak. 
It is interesting to see that the zero point shift to the 
longer (shorter) $L$ by about 200km as the average density 
becomes smaller (larger).

\vskip 2mm
\noindent
(d) The effect from the uncertainty of matter fluctuations 

As we discussed in the subsection (b), the main contribution 
from matter fluctuations is from the term containing the 
Fourier coefficient $a_2$ for symmetric fluctuations, 
and similarly $a_1$ for asymmetric fluctuations. 
Since  $a_2$ is determined from the most dense matter part 
(middle part) along the neutrino path. On the other hand, 
higher modes $a_{2n}$ ($n=2,3,\cdots$) are determined mainly 
by the crust of earth. Therefore, for $L>6000$km, $a_2$ is 
considered to be rather unambiguously determined reflecting 
the deep inside structure of mantle. 

For asymmetric fluctuations, $a_1$ reflects the global 
asymmetric feature of matter profile. For distances 
at a neutrino factory, neutrinos pass mainly through the 
mantle and we do not expect much uncertainty $a_1$. 
For the shorter length, neutrinos pass through the crust 
and sometimes the sea. In this situation, the matter 
profile with large asymmetric matter fluctuations may 
need to be considered.[9] 

For the sake of argument, we assume that there is about 10\% 
uncertainty for $a_2$, 
i.e., $a_{2n}=(1\pm 0.1)(a_{2n})_{PREM}$. Since 
$\Delta P_{\nu_\alpha \nu_\beta}^{T(2s)}$ 
depends linearly on $a_2$, the 10\% uncertainty for $a_2$ 
gives the same uncertainty for 
$\Delta P_{\nu_\alpha \nu_\beta}^{T(2s)}$. For distances 
where the 2nd order term is neglected, the uncertainty 
from symmetric fluctuations is negligible ($L<$6000km). 
In distances where the 2nd order term dominates, then 
10\% uncertainty appears ($L>10000$km). The uncertainty 
for $L>8000$km is smaller than 10\%. For 7000km$<L<$8000km, 
the uncertainty is larger than 10\%. In Fig.10, we 
show how T violation becomes uncertain if $a_2$ has 10\% 
uncertainty. 

For asymmetric fluctuations, we assumed that 
$a_{2n-1}=0.1(a_{2n})_{PREM}$ in addition to 
symmetric matter fluctuations determined from PREM.  
The effect from asymmetric fluctuations is 
very similar to the case of symmetric fluctuations 
and it is shown in Fig.10.

\section{Summary}

In this paper, we derived the analytic formula for 
T violation up to the 3rd order term for small quantities, 
$\Delta m_{21}^2L/2E$ and $\delta a(x)L/2E$. By using 
this formula, we examined the $E$ and $L$ dependence of 
T violation. 

We showed that the effect from both symmetric and 
asymmetric matter fluctuations are negligible 
for $L<$6000km and the constant (average) matter 
approximation (the 1st order term) is valid. Therefore, 
T violation 
contains the uncertainty from the average matter 
aside from mixing angles. Since the average is 
considered to be determined with much less uncertainty 
than matter fluctuations, T violation is determined 
with a good accuracy for $L<$6000km. 

For $L>6000$km, the situation changes. Matter 
fluctuations (the 2nd order term) give a sizable effect 
to T violation. Moreover, the 1st and the 2nd order 
terms contribute destructively for 7000km$<L<$7700km. 
As a result, the T violation becomes very small. 

In Fig.7, we showed the $L$ dependence of T violation at 
the lower energy peak (the peak at $E=$5.5GeV for $L=$6000km) 
and the higher energy peak 
(the peak for $E=$10GeV for $L=$6000km). T violation at 
the lower energy peak has the 
largest value for 5000km$<L<$6000km and $L\sim$10000km 
and T violation 
at the higher energy peak does for 3000km$<L<$4000km. 
T violation becomes zero at around $L=$7300km. 
This is due to the resonance effect, which 
we explained why this happens by using the analytic formula. 
For $E=30$GeV, T violation behaves similarly and 
has the largest value at 5000km. 

We also examined how the $L$ dependence of T violation 
varies as the average matter is changed by $\pm 5\%$, 
although we believe that the average matter density 
is determined 
with much less uncertainty. We found that the distance which 
gives zero of T violation shifts about $\mp$200km.

\vskip 5mm
{\Huge Acknowledgment}
This work is supported in part by 
the Japanese Grant-in-Aid for Scientific Research of 
Ministry of Education, Culture, Sports, Science and 
Technology, No.12047218.

\newpage
\setcounter{section}{0}
\renewcommand{\thesection}{\Alph{section}}
\renewcommand{\theequation}{\thesection .\arabic{equation}}
\newcommand{\apsc}[1]{\stepcounter{section}\noindent
\setcounter{equation}{0}{\Large{\bf{Appendix\,\thesection\,:\,{#1}}}}}

\apsc{The brief summary of our previous result}

In Ref.1, we derived T violation fromula up to the 2nd order 
with respect to small quantities of $\Delta m_{21}^2 L/4E$ 
and $\delta a L/4E$, where $\delta a$ represents matter 
fluctuations, where both symmetric $\delta a_s$ and 
asymmetric $\delta a_a$ fluctuations are taken into account. 
Here, we give the brief summary of the previous result and 
the 3rd order calculation needed to compute the next order 
symmetric matter fluctuation effect. 

We express the S-matrix as
\bea
S=S_{00}+\sum_{n,m=0,1,2,\cdots} S_{01,1}^{(n,m)}\;,
\ena
where $S_{00}$ is the 0th order term from $H_{00}$, 
$S_{01,1}^{(n,m)}$ represents the (n+m)th order term 
from $H_{01}^n\times H_1^m$, i.e., the term of order 
$(\Delta m_{21}^2 L/4E)^n (\delta a L/4E)^m$. 

The contributions to T violation  is summarized as follows: 
(1) The 1st order term is from $S_{00}S_{01,1}^{(1,0)*}$ 
and is proportional to $\sin \delta$ and contains 
the contribution from the constant (average) matter. 
(2) The 2nd order term is from $S_{00}S_{01,1}^{(1,1)*}$ 
and $S_{01}S_{01,1}^{(0,1)*}$. The symmetric and asymmetric 
matter fluctuations contribute to $\sin \delta$ and 
$\cos \delta$ terms, respectively. 
(3) The 2nd order term, $S_{00}S_{01,1}^{(0,2)*}$ and the 
3rd order term, $S_{00}S_{01,1}^{(0,3)*}$ do not contribute 
to T violation. 

Firstly, we present the result given in Ref.1:
\bea
&&S_{00}=e^{-iH_{00}L}=\tilde{U}_0 P(L)
\tilde{U}_0^\dagger\nonumber\\
&=&\frac12
\pmatrix{\phi_+-c_{2\tp}\phi_-
& s_{y} s_{2\tp} e^{-i\delta}\phi_- 
& c_{y} s_{2\tp} e^{-i\delta}\phi_- \cr
  s_{y} s_{2\tp} e^{i\delta}\phi_- 
& \phi_++s_y^2c_{2\tp}\phi_--c_y^2(\phi_{2-}-\phi_{1-})
& s_{2y}
\left(c_{\tp}^2 \phi_{2-}-s_{\tp}^2 \phi_{1-}\right)\cr
  c_{y} s_{2\tp} e^{i\delta}\phi_- 
& s_{2y}\left(c_{\tp}^2 \phi_{2-}-s_{\tp}^2 \phi_{1-}
\right)
& \phi_++c_y^2c_{2\tp}\phi_--s_y^2(\phi_{2-}-\phi_{1-})\cr
}\;,
\nonumber\\
\ena
where $\phi_\pm$ and $\phi_{i\pm}$ ($i=1,2$) are defined as follows.
\bea
\phi_{\pm}&=& e^{-ia_+ L}\pm e^{-ia_- L}\;,\nonumber\\
\phi_{1\pm}&=&e^{-ia_0 L}\pm e^{-ia_- L}\;,
\;\; \phi_{2\pm}=e^{-ia_+ L}\pm e^{-ia_0 L}\;.
\ena
$S_{01,1}^{(n,m)}(n\ge 1)$ is parametrized by  
\bea
S_{01,1}^{(n,m)}&=&\tilde{U}_0 
\pmatrix{0&{\cal A}^{(n,m)}&0\cr {\cal A'}^{(n,m)}&0&
{\cal B}^{(n,m)}\cr 0&{\cal B'}^{(n,m)}&0\cr}
\tilde{U}_0^\dagger
\nonumber\\
&=&\pmatrix{0&c_y {\cal P}^{(n,m)}&-s_y {\cal P}^{(n,m)}\cr
c_y {\cal P'}^{(n,m)} 
&-s_yc_y(e^{i\delta} {\cal Q}^{(n,m)}
+e^{-i\delta} {\cal Q'}^{(n,m)}) 
&e^{i\delta} s_y^2{\cal Q}^{(n,m)}
-e^{-i\delta} c_y^2{\cal Q'}^{(n,m)} \cr
-s_y {\cal P'}^{(n,m)}   
&-e^{i\delta} c_y^2{\cal Q}^{(n,m)}
+e^{-i\delta} s_y^2{\cal Q'}^{(n,m)} 
& s_y c_y (e^{i\delta} {\cal Q}^{(n,m)} 
+ e^{-i\delta} {\cal Q'}^{(n,m)})
\cr}\;,
\nonumber\\
\ena
where
\bea
{\cal P}^{(n,m)}&=&(c_\tp {\cal A}^{(n,m)}+s_\tp {\cal
B}^{(n,m)})\;,\; 
{\cal Q}^{(n,m)}=(s_\tp {\cal A}^{(n,m)}-c_\tp {\cal B}^{(n,m)})\;,
\nonumber\\
{\cal P'}^{(n,m)}&=&(c_\tp {\cal A'}^{(n,m)}+s_\tp {\cal
B'}^{(n,m)})\;,\; 
{\cal Q'}^{(n,m)}=(s_\tp {\cal A'}^{(n,m)}-c_\tp {\cal B'}^{(n,m)})\;.
\ena

The $S_{01,1}^{(1,0)}$ is obatained from 
\bea
{\cal A}^{(1,0)}&=& {\cal A'}^{(1,0)}
=\frac{\Delta m^2_{21}}{4E}\frac{s_{2x}c_{z-\tp}}
{k_1}\phi_{1-}\;,\nonumber\\
{\cal B}^{(1,0)}&=& {\cal B'}^{(1,0)}=
-\frac{\Delta m^2_{21}}{4E}\frac{s_{2x}s_{z-\tp}}
{k_2}\phi_{2-}\;.
\ena
The $S_{01,1}^{(1,1)}$ is obatained from 
\bea
{\cal A}^{(1,1s)}&=&{\cal A'}^{(1,1s)}
\nonumber\\
&=&\frac{\Delta m_{21}^2 s_{2x}}{4E} c_\tp \left\{
\left(\frac{c_\tp c_{z-\tp}}{k_1}+\frac{s_\tp s_{z-\tp}}{k_2}\right) 
\right. \nonumber\\
&&\times \left. 
\left(\sum_{n=1,2,\cdots}\frac{a_{2n}k_1}{E(k_1^2-q_{2n}^2)}\right)
\phi_{1-}
-\frac{s_\tp s_{z-\tp}}{k_2}\sum_{n=1,2,\cdots}
\left(\frac{a_{2n}k}{E(k^2-q_{2n}^2)}\right)\phi_{-}\right\}\;,
\nonumber\\
{\cal B}^{(1,1s)}&=&{\cal B'}^{(1,1s)}
\nonumber\\ &=&
\frac{\Delta m_{21}^2s_{2x}}{4E}s_\tp\left\{
\left(\frac{c_\tp c_{z-\tp}}{k_1}+\frac{s_\tp s_{z-\tp}}{k_2}\right) 
\right. \nonumber\\
&&\times \left.
\left(\sum_{n=1,2,\cdots}\frac{a_{2n}k_2}{E(k_2^2-q_{2n}^2)}\right)
\phi_{2-}
-\frac{c_\tp c_{z-\tp}}{k_1}\left(\sum_{n=1,2,\cdots}
\frac{a_{2n}k}{E(k^2-q_{2n}^2)}\right)\phi_{-}\right\}\;,
\ena
for symmetric matter fluctuations and 
\bea
{\cal A}^{(1,1a)}&=&-{\cal A'}^{(1,1a)}
\nonumber\\
&=&-\frac{\Delta m_{21}^2 s_{2x}}{4E} c_\tp \left\{
\left(\frac{c_\tp c_{z-\tp}}{k_1}+\frac{s_\tp s_{z-\tp}}{k_2}\right) 
\right. \nonumber\\
&&\times \left. 
\left(\sum_{n=0,1,\cdots}\frac{a_{2n+1}k_1}{E(k_1^2-q_{2n+1}^2)}\right)
\phi_{1-}
-\frac{s_\tp s_{z-\tp}}{k_2}\sum_{n=0,1,\cdots}
\left(\frac{a_{2n+1}k}{E(k^2-q_{2n+1}^2)}\right)\phi_{-}\right\}\;, 
\nonumber\\
{\cal B}^{(1,1a)}&=&-{\cal B'}^{(1,1a)}\nonumber\\
&=&\frac{\Delta m_{21}^2s_{2x}}{4E}s_\tp\left\{
\left(\frac{c_\tp c_{z-\tp}}{k_1}+\frac{s_\tp s_{z-\tp}}{k_2}\right) 
\right. \nonumber\\
&&\times \left.
\left(\sum_{n=0,1,\cdots}\frac{a_{2n+1}k_2}{E(k_2^2-q_{2n+1}^2)}\right)
\phi_{2-}
-\frac{c_\tp c_{z-\tp}}{k_1}\left(\sum_{n=0,1,\cdots}
\frac{a_{2n+1}k}{E(k^2-q_{2n+1}^2)}\right)\phi_{-}\right\}\;,
\ena
for asymmetric fluctuations. 

$S_{01,1}^{(0,n)}$ is given by 
\bea
S_{01,1}^{(0,n)}=
\tilde{U}_0\pmatrix{{\cal E}^{(0,n)}&0&{\cal C}^{(0,n)}\cr
0&0&0\cr {\cal D}^{(0,n)}&0&{\cal F}^{(0,n)}}\tilde{U}_0^\dagger
=\frac12
\pmatrix{\alpha_{n}^{(+)}&e^{-i\delta}s_y\beta_{n}^{(+)}&
e^{-i\delta}c_y \beta_{n}^{(+)}\cr
e^{i\delta}s_y\beta_{n}^{(-)}&s_y^2\alpha_{n}^{(-)}
&s_yc_y\alpha_{n}^{(-)}\cr
e^{i\delta}c_y \beta_{n}^{(-)}&s_yc_y\alpha_{n}^{(-)}
&c_y^2\alpha_{n}^{(-)}\cr}\;,
\ena
where
\bea
\alpha_{n}^{(\pm)}&=&{\cal E}^{(0,n)}+{\cal F}^{(0,n)}
\pm \left(c_{2\tp}({\cal E}^{(0,n)}-{\cal F}^{(0,n)})+
s_{2\tp}({\cal C}^{(0,n)}+{\cal D}^{(0,n)})\right)  \;,
\nonumber\\
\beta_{n}^{(\pm)}&=&-s_{2\tp}({\cal E}^{(0,n)}-{\cal F}^{(0,n)})
+c_{2\tp}({\cal C}^{(0,n)}+{\cal D}^{(0,n)})\pm 
({\cal C}^{(0,n)}-{\cal D}^{(0,n)}) \;.
\ena

$S_{01,1}^{(0,1)}$ is derived from 
\bea
{\cal C}^{(0,1)}+{\cal D}^{(0,1)}
&=&\frac{s_{2\tp}}{2E}\left(
  \sum_{n\neq 0}\frac{a_{2n}}{k+q_{2n}}\right)\phi_-\;,
  \nonumber\\
{\cal C}^{(0,1)}-{\cal D}^{(0,1)}&=&
-\frac{s_{2\tp}}{2E}\left(
\sum_{n}\frac{a_{2n+1}}{k+q_{2n+1}}\right)\phi_+\;,
\nonumber\\
{\cal E}^{(0,1)}&=&{\cal F}^{(0,1)}=0\;.
\ena

By using the above fromula, we obtained the 2nd order 
formula for T violation. 

Finally, we give the result 
for $S_{01,1}^{(0,2)}$ which is needed to estimate 
the 3rd order correction. $S_{01,1}^{(0,2)}$ is 
calculated from
\bea
{\cal C}^{(0,2)}+{\cal D}^{(0,2)}&=&
\frac{s_{2\tp}c_{2\tp}}{(2E)^2}\left(
\sum_{m+n={\rm even}}\frac{a_na_m}{(k+q_n)(k+q_n+q_m)}\right)
\phi_-\;,
\nonumber\\
{\cal C}^{(0,2)}-{\cal D}^{(0,2)}&=&
-\frac{s_{2\tp}c_{2\tp}}{(2E)^2}\left(
\sum_{m+n={\rm odd}}\frac{a_na_m}{(k+q_n)(k+q_n+q_m)}\right)\phi_+
\;,
\nonumber\\
{\cal E}^{(0,2)}-{\cal F}^{(0,2)}&=&
\frac{s_{2\tp}^2}{8(2E)^2}\left\{ \left(
\sum{n,m}
 \frac{((-1)^{n}+1)((-1)^m+1)a_n a_m }{(k+q_n)(k+q_m)}
\right)\phi_- \right.\nonumber\\
&&\hskip 1.5cm +\left. 2L\left(\sum_{n}\frac{a_na_{-n}}{k+q_n}
\right)i\phi_+
\right\}\;.
\ena

\vskip 5mm
\apsc{The 3rd order calculation}

We consider the 3rd order correction form $H_{01}H_1^2$ where 
we take symmetric matter fluctuations for $H_1$. The contribution 
arises from $S_{00}S_{01,1}^{(1,2)*}$, 
$S_{01,1}^{(1,0)}S_{01,1}^{(0,2)*}$ and 
$S_{01,1}^{(1,1)}S_{01,1}^{(0,1)*}$. The latter two can be calculated 
the result given in Appendix A. Here, we give the result for 
$S_{01,1}^{(1,2)}$ which is given from 
\bea
{\cal A}^{(1,2s)}&=&{\cal A'}^{(1,2s)}
=\frac{\Delta m_{21}^2 s_{2x}}{4E} c_\tp
\nonumber\\ 
&\times&\left\{
\left(\frac{c_\tp c_{z-\tp}}{k_1}+\frac{s_\tp s_{z-\tp}}{k_2}
\right)
\left(\sum_{n\neq 0}\frac{a_{2n}a_{2m}}{(2E)^2}
\left(
\frac{c_{\tp}^2}{k_1+q_{2m}} -\frac{s_{\tp}^2}{k_2-q_{2m}}\right)
\frac1{k_1+q_{2n+2m}} 
\right)
\phi_{1-}\right.
\nonumber\\
&&\left. -\frac{c_\tp c_{z-\tp}}{k_1}s_\tp^2L
\left(
\sum_{n\neq 0}\frac{a_{2n}a_{-2n}}{(2E)^2}
\frac1{k+q_{2n}} 
\right) e^{-ia_-L}
\right\}\;,
\nonumber\\
{\cal B}^{(1,2s)}&=&{\cal B'}^{(1,2s)}
=-\frac{\Delta m_{21}^2s_{2x}}{4E}s_\tp
\nonumber\\
&\times&\left\{
\left(\frac{c_\tp c_{z-\tp}}{k_1}
+\frac{s_\tp s_{z-\tp}}{k_2}\right) 
\left(\sum_{n\neq 0}\frac{a_{2n}a_{2m}}{(2E)^2}
\left(\frac{c_{\tp}^2}{k_1+q_{2m}} -\frac{s_{\tp}^2}{k_2-q_{2m}}
\right)\frac1{k_2-q_{2n+2m}} \right)\phi_{1-}
\right.\nonumber\\
&&\left. -\frac{s_\tp s_{z-\tp}}{k_2}c_\tp^2L
\left(\sum_{n\neq 0}\frac{a_{2n}a_{-2n}}{(2E)^2}
\frac1{k+q_{2n}} \right) e^{-ia_+L}\right\}\;.
\nonumber\\
\ena
By using the above formula, we can derive the 3rd order 
contribution to T violation.

\begin{figure}[pht]
\epsfxsize=13cm
\centerline{\epsfbox{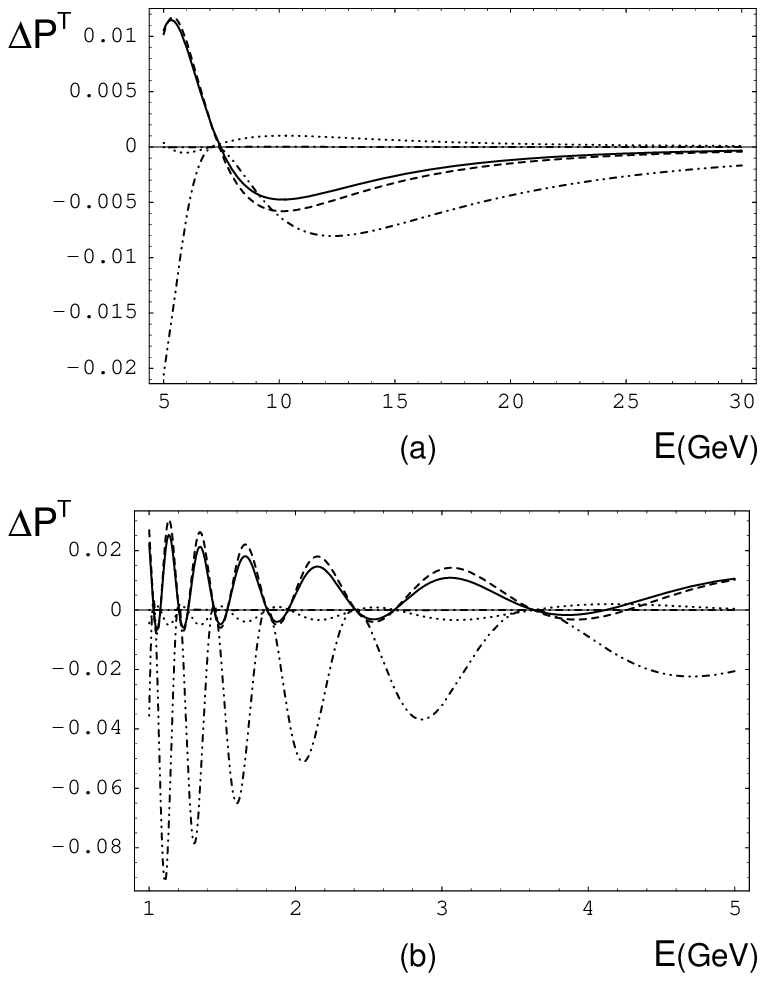}}
\caption{The energy dependence of the T violation, 
$\Delta P^T_{\nu_e \nu_\mu}$ 
with $L=6000$km for (a) 5GeV$<E<$30GeV and (b) 1GeV$<E<$5GeV. 
The dashed line, the dotted line and the dash-dotted line
show the 1st, the 2nd and the 3rd order terms. 
The solid line represents the sum of them. 
The vacuum (no matter) case is shown by the 
dash-twodotted line for the comparison.
In this plot, we use $\sin 2\theta_x=\sin 2\theta_y=1$, 
$\sin \theta_z=0.1$,  $\Delta m_{21}^2=5\cdot 10^{-5}{\rm eV}^2$,  
$\Delta m_{31}^2=3\cdot 10^{-3}{\rm eV}^2$ and $\delta=\pi/4$.}
\end{figure}

\begin{figure}[pht]
\epsfxsize=13cm
\centerline{\epsfbox{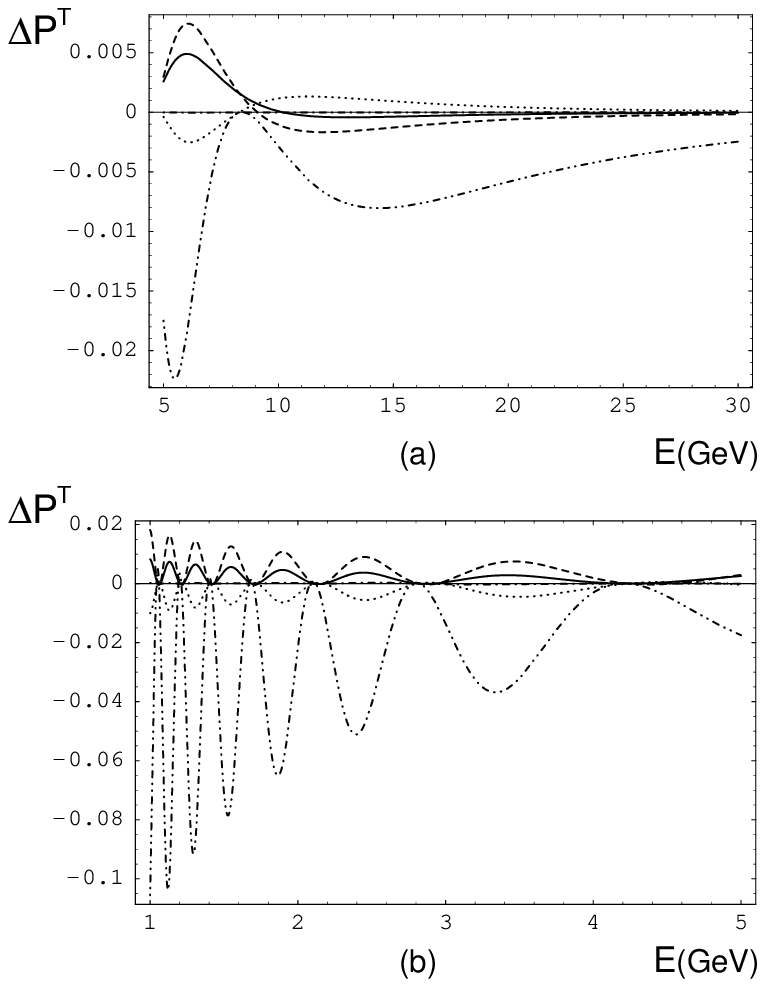}}
\caption{The energy dependence of the T violation, 
$\Delta P^T_{\nu_e \nu_\mu}$ 
with $L=7000$km for (a) 5GeV$<E<$30GeV and (b) 1GeV$<E<$5GeV. 
The species of the lines and the oscillation parameters 
are the same as those in Fig.1.}
\end{figure}

\begin{figure}[pht]
\epsfxsize=13cm
\centerline{\epsfbox{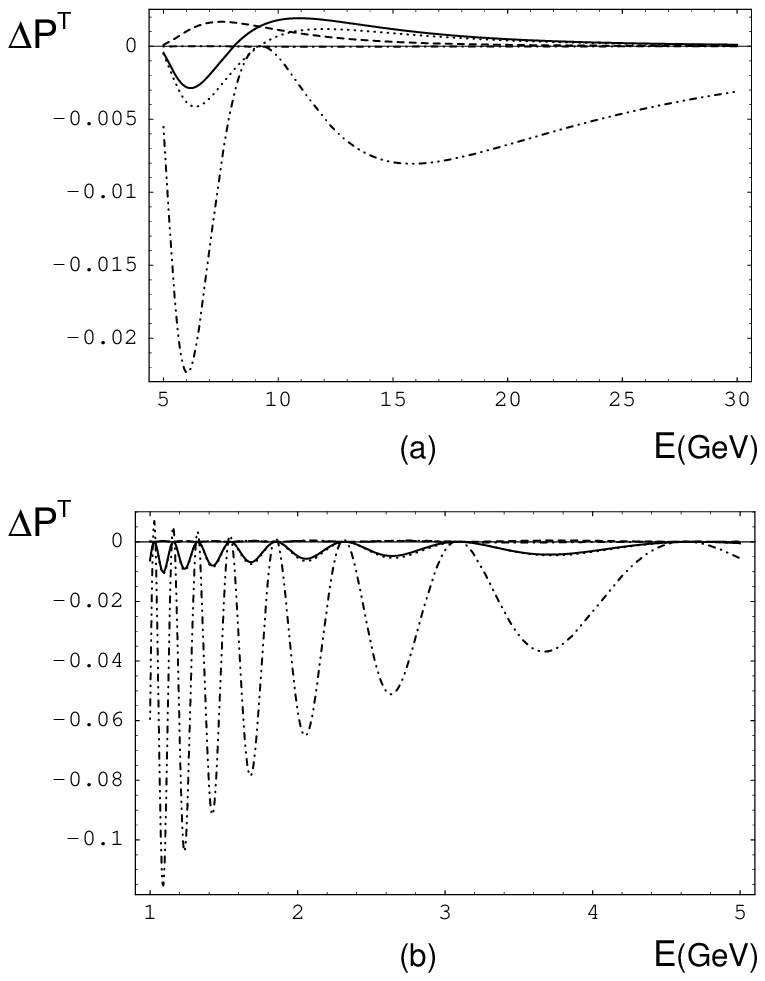}}
\caption{The energy dependence of the T violation, 
$\Delta P^T_{\nu_e \nu_\mu}$  
with $L=7700$km for (a) 5GeV$<E<$30GeV and (b) 1GeV$<E<$5GeV. 
The species of the lines and the oscillation parameters 
are the same as those in Fig.1.}
\end{figure}

\begin{figure}[pht]
\epsfxsize=13cm
\centerline{\epsfbox{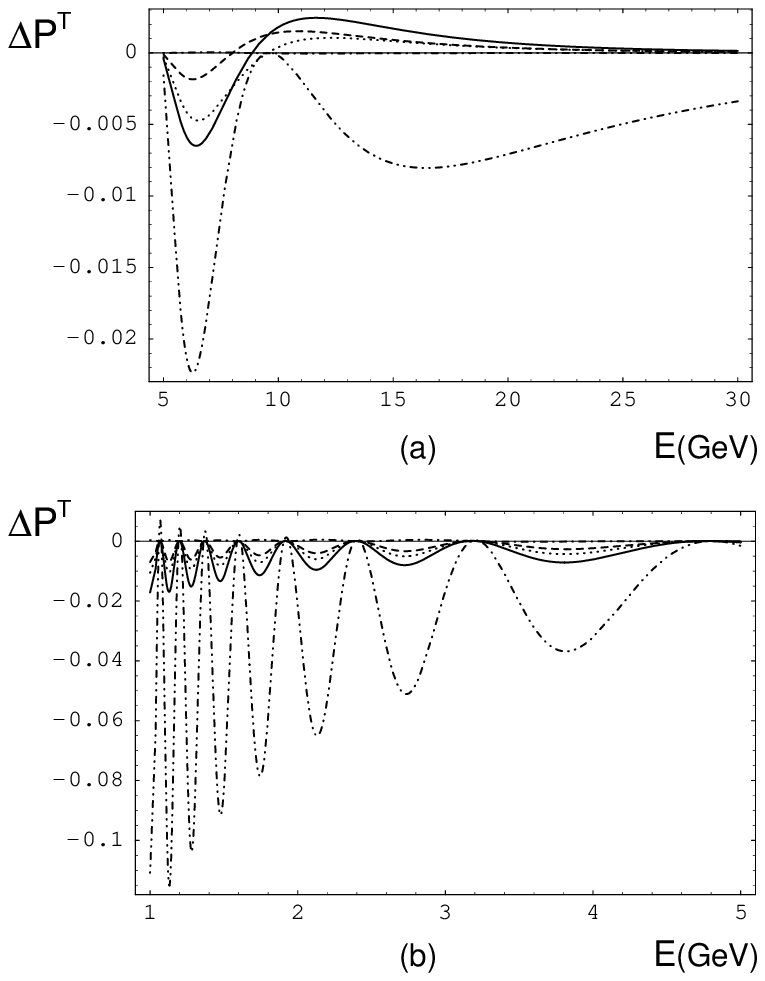}}
\caption{The energy dependence of the T violation, 
$\Delta P^T_{\nu_e \nu_\mu}$ 
with $L=8000$km for (a) 5GeV$<E<$30GeV and (b) 1GeV$<E<$5GeV. 
The species of the lines and the oscillation parameters 
are the same as those in Fig.1.}
\end{figure}

\begin{figure}[pht]
\epsfxsize=13cm
\centerline{\epsfbox{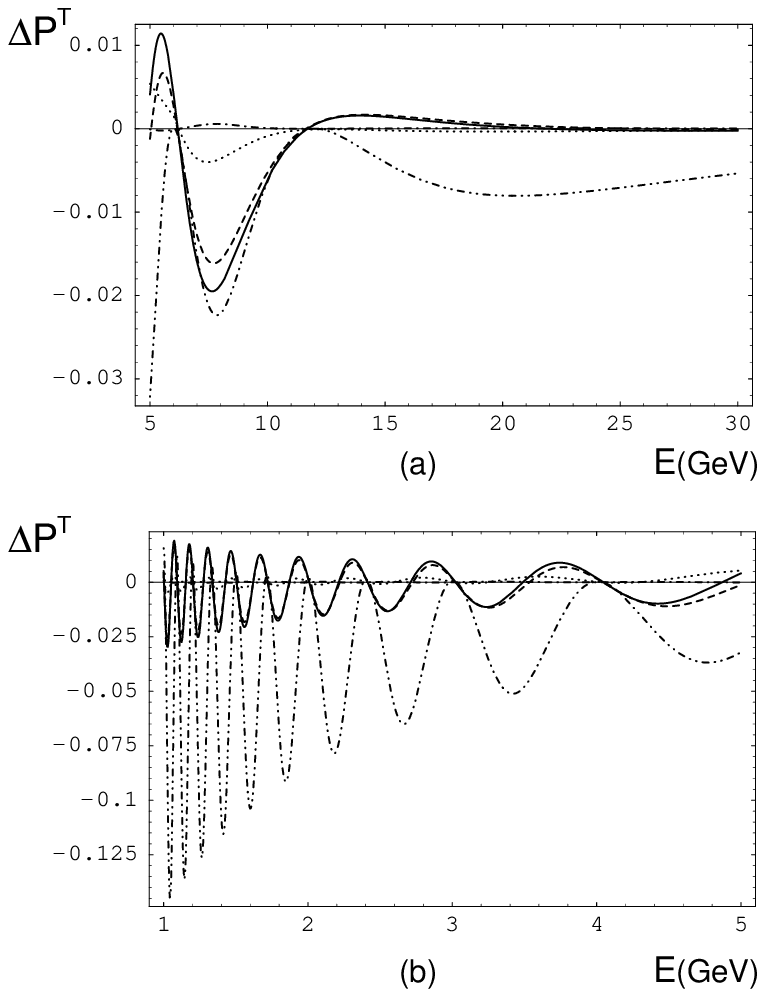}}
\caption{The energy dependence of the T violation, 
$\Delta P^T_{\nu_e \nu_\mu}$  
with $L=10000$km for (a) 5GeV$<E<$30GeV and (b) 1GeV$<E<$5GeV. 
The species of the lines and the oscillation parameters 
for (a) 5GeV$<E<$30GeV and (b) 1GeV$<E<$5GeVare the same as those in
Fig.1.}
\end{figure}

\begin{figure}[pht]
\epsfxsize=13cm
\centerline{\epsfbox{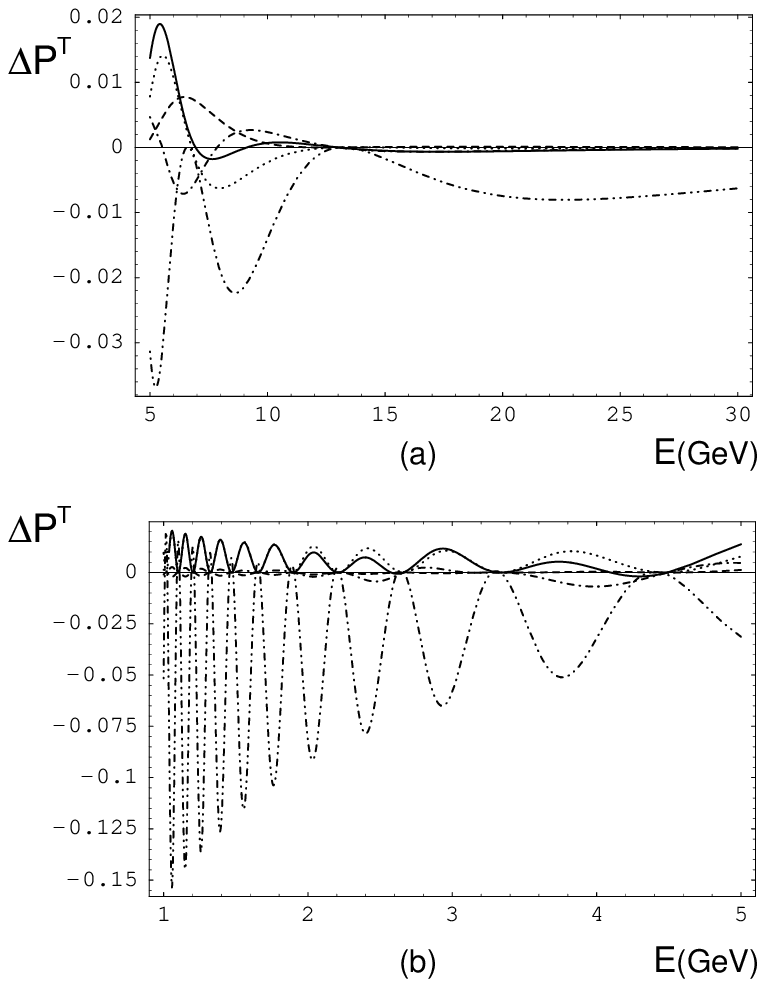}}
\caption{The energy dependence of the T violation,  
$\Delta P^T_{\nu_e \nu_\mu}$ 
with $L=11000$km for (a) 5GeV$<E<$30GeV and (b) 1GeV$<E<$5GeV. 
The species of the lines and the oscillation parameters 
are the same as those in Fig.1.}
\end{figure}

\begin{figure}[pht]
\epsfxsize=11cm
\centerline{\epsfbox{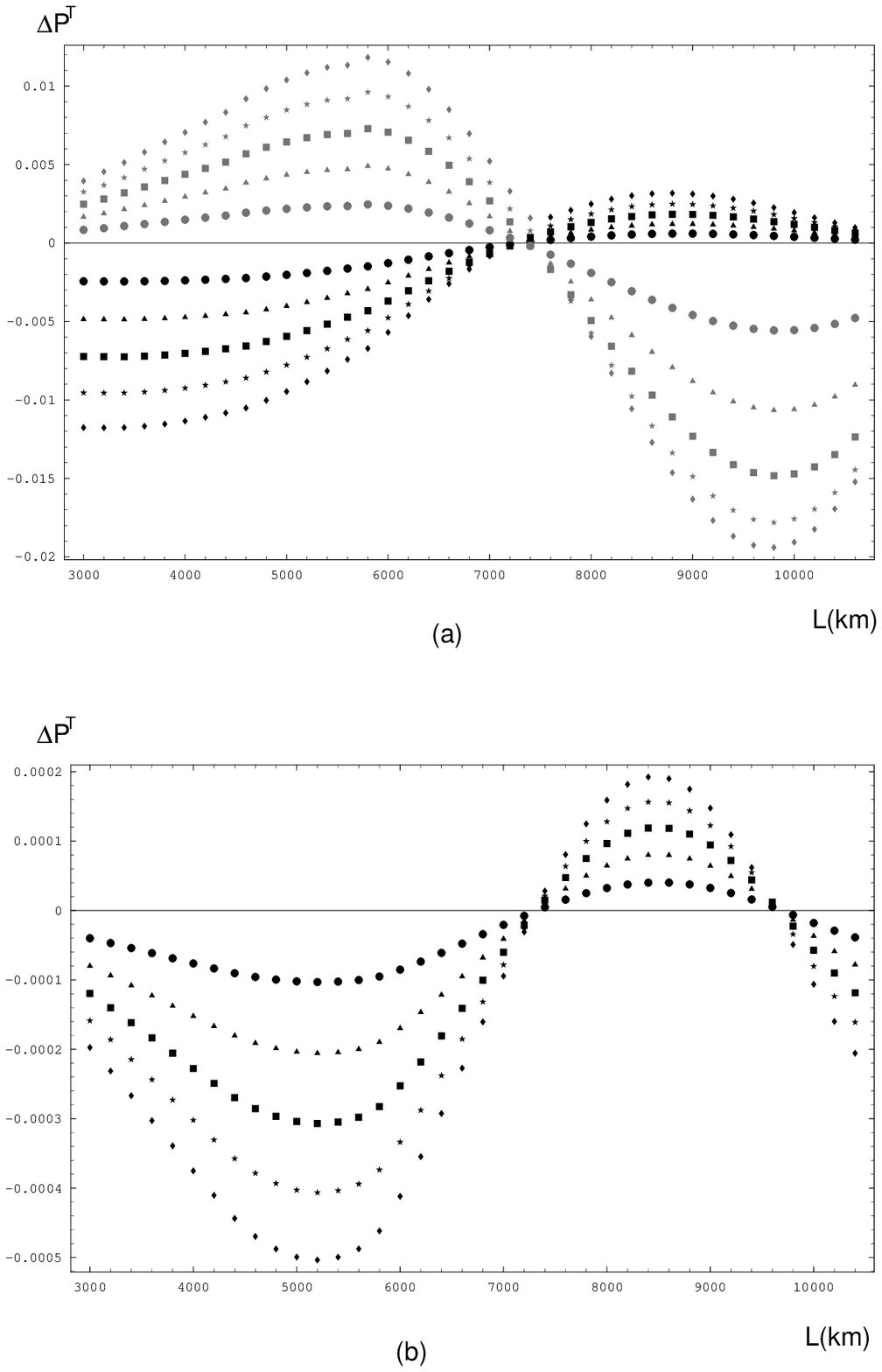}}
\caption{The $L$ dependence of T violation,
$\Delta^T_{\nu_e\nu_{\mu}}$.
In Fig.(a) T violation at the lower energy peak ($E=5\sim 8$ GeV)
which is shown by faint points and the higher energy peak ($E=10\sim
13$ GeV)
which is shown by dark points.
Diamonds, stars, boxes, triangles and circles show T violation for 
$s_z=\sin\theta_{13}=0.1$, $0.08$, $0.06$, $0.04$ and $0.02$
respectively.
In Fig. (b) we show the values of T violation $\Delta
P^T_{\nu_e\nu_\mu}$ 
at $E=30$ GeV. The oscillation parameters except for $s_z$ are the
same as those in Fig.1.
}
\end{figure}

\begin{figure}[pht]
\epsfxsize=15cm
\centerline{\epsfbox{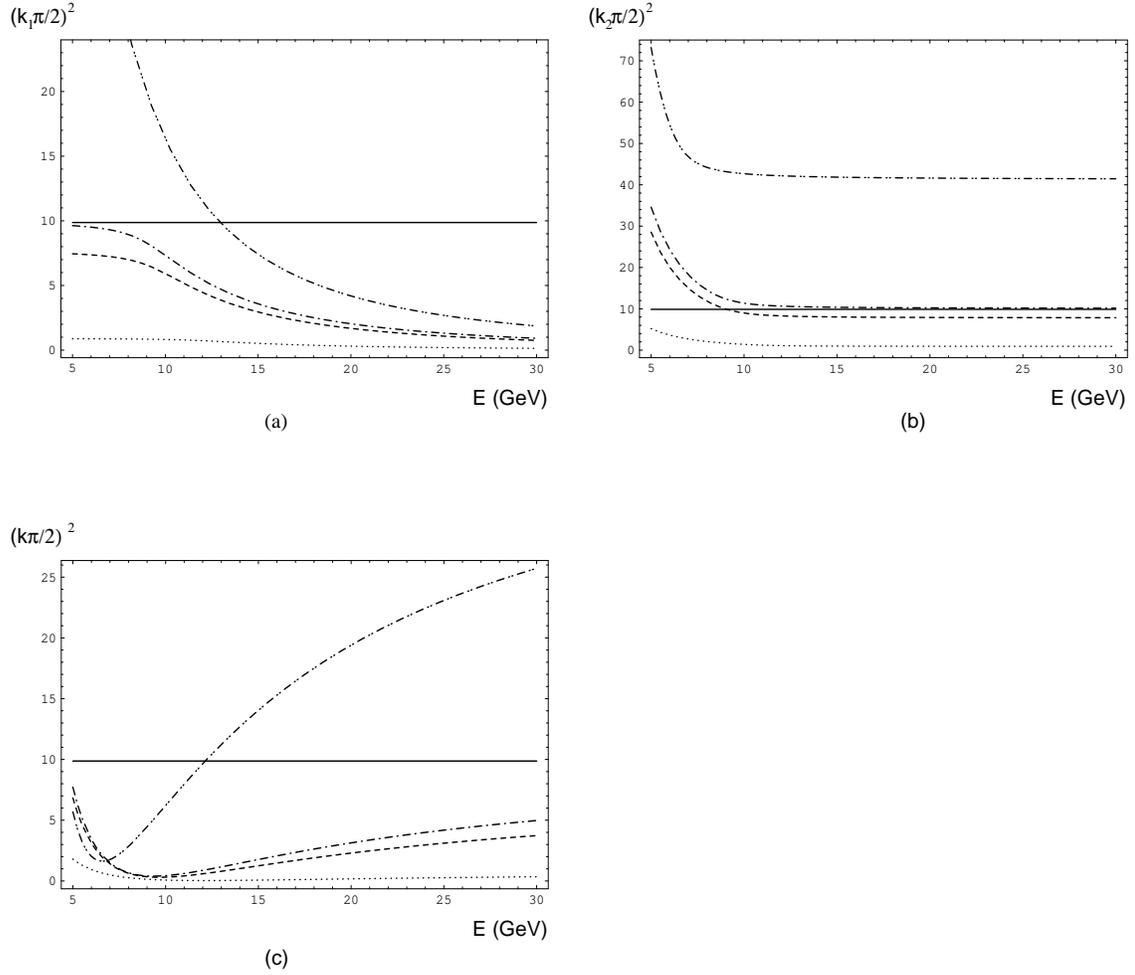}}
\caption{The energy dependence of $(k_1L/2)^2$(Fig.(a)),
 $(k_2L/2)^2$(Fig.(b)) and $(kL/2)^2$(Fig.(c)). Dotted,
dashed, dash-dotted and dash-twodotted lines correspond to $L=3000,
7000,
7700, 11000$ km, respectively. The solid line shows $\pi^2$.
The oscillation parameters are the same as those in Fig.1.}
\end{figure}

\begin{figure}[pht]
\epsfxsize=15cm
\centerline{\epsfbox{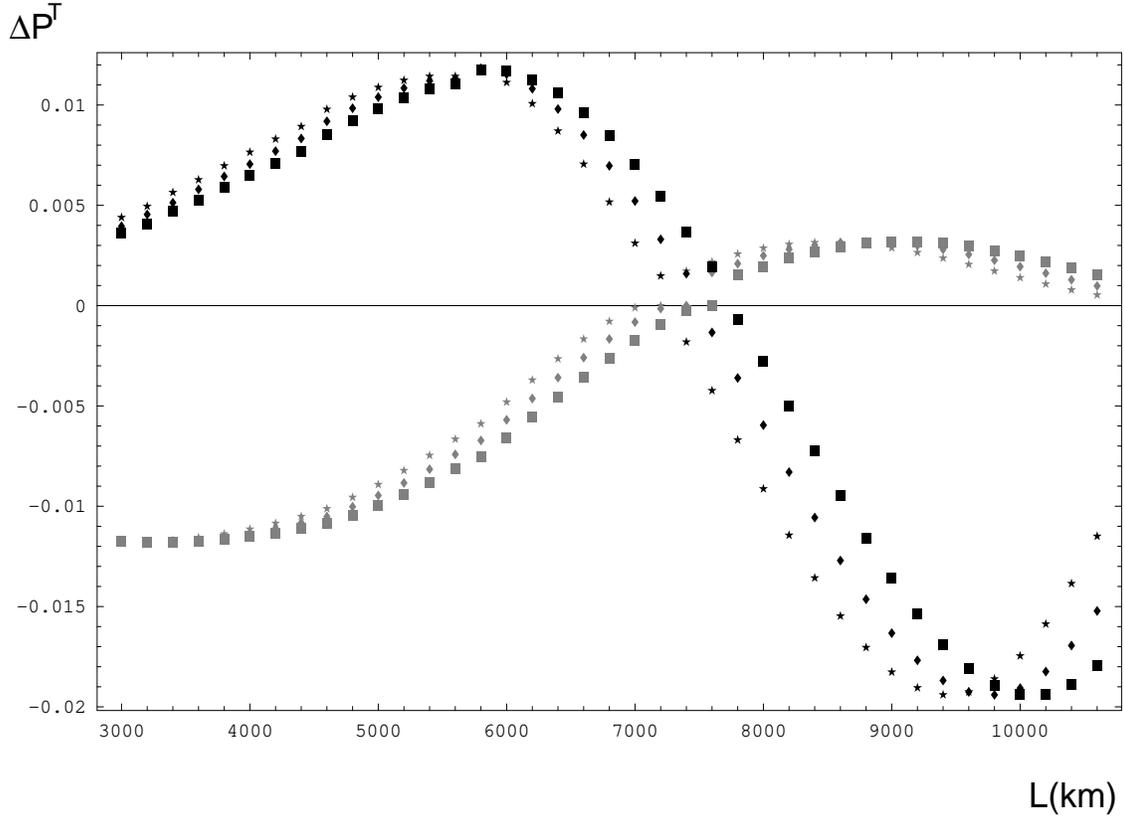}}
\caption{The average density ($\bar{a}$) dependence of T violation, 
$\Delta P^T_{\nu_e\nu_{\mu}}$.
Diamonds, boxes and stars correspond to the average density value from
PREM, $5\%$ smaller value and $5\%$ larger value. Faint and dark
points 
correspond to the lower energy peak ($E=5\sim 8$ GeV) and 
the higher energy peak ($E=10\sim 13$ GeV). The oscillation parameters 
are the same as those in Fig.1.}
\end{figure}

\begin{figure}[pht]
\epsfxsize=13cm
\centerline{\epsfbox{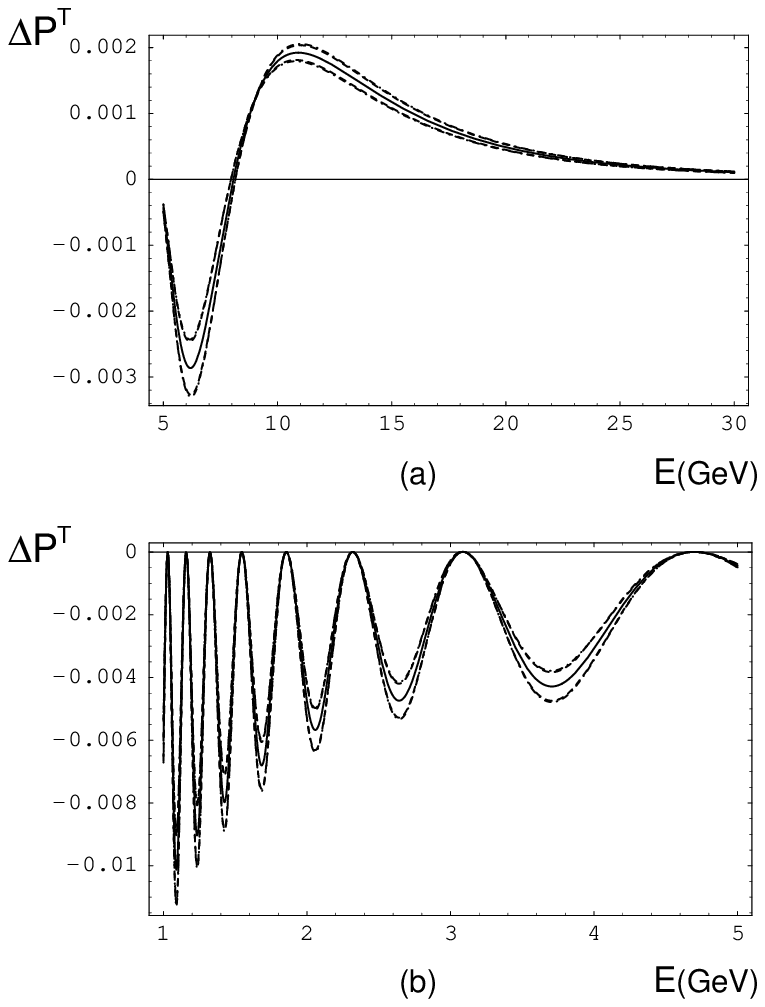}}
\caption{
The variation of T violation, $\Delta P^T_{\nu_e \nu_\mu}$ 
by the deformation of the shape of matter profile for $L=7700$km.
Dotted lines show changes of T violation 
when Fourier coefficients for symmetric matter fluctuations $a_{2n}$ 
are varied by 10$\%$ from PREM.
Dash-dotted lines are those when
asymmetric matter fluctuations 
are associated to be about 10$\%$ of PREM symmetric 
matter fluctuations.
The oscillation parameters 
are the same as those in Fig.1.}
\end{figure}

\end{document}